\documentclass[10pt, final, conference, letterpaper]{IEEEtran}
\usepackage[left=.6in,right=.6in,top=.719in,bottom=.99in]{geometry}
\usepackage{amsmath,amssymb,dsfont,stfloats,color,url,bbm}
\usepackage[pdftex]{graphicx}
\usepackage[caption=false,font=footnotesize]{subfig}
\usepackage{mathabx}

\DeclareGraphicsExtensions{.eps,.pdf,.png,.jpg,.gif,.jpeg,.pstex}

\usepackage[noadjust]{cite}
\usepackage{url}

\newcommand*{\Resize}[2]{\resizebox{#1}{!}{$#2$}}%

\newfont{\bbb}{msbm10 scaled 700}

\newfont{\bb}{msbm10 scaled 1100}
\newcommand{\CC}{\mbox{\bb C}}
\newcommand{\PP}{\mbox{\bb P}}

\newcommand{\EE}{\mbox{\bb E}}

\newcommand{\HH}{\mbox{\bb H}}
\newcommand{\SSS}{\mbox{\bb S}}
\newcommand{\UU}{\mbox{\bb U}}

\newcommand{\BB}{\mbox{\bb B}}
\renewcommand{\AA}{\mbox{\bb A}}

\newcommand{\yy}{\mathbbm{y}}
\newcommand{\xx}{\mathbbm{x}}
\newcommand{\zz}{\mathbbm{z}}
\newcommand{\sss}{\mathbbm{s}}

\newcommand{\hh}{\mathbbm{h}}
\newcommand{\uu}{\mathbbm{u}}
\newcommand{\vvv}{\mathbbm{v}}

\newcommand{\hv}{{\bf h}}

\newcommand{\qv}{{\bf q}}

\newcommand{\uv}{{\bf u}}

\newcommand{\vv}{{\bf v}}

\newcommand{\yv}{{\bf y}}
\newcommand{\zv}{{\bf z}}
\newcommand{\zerov}{{\bf 0}}

\newcommand{\Fm}{{\bf F}}

\newcommand{\Hm}{{\bf H}}
\newcommand{\Id}{{\bf I}}

\newcommand{\Ym}{{\bf Y}}
\newcommand{\Zm}{{\bf Z}}

\newcommand{\Cc}{{\cal C}}

\newcommand{\Ec}{{\cal E}}

\newcommand{\Gc}{{\cal G}}

\newcommand{\Nc}{{\cal N}}

\newcommand{\Sc}{{\cal S}}

\newcommand{\Uc}{{\cal U}}

\newcommand{\nuv}{\hbox{\boldmath$\nu$}}
\newcommand{\muv}{\hbox{\boldmath$\mu$}}

\newcommand{\phiv}{\hbox{\boldmath$\phi$}}

\newcommand{\Thetam}{\hbox{\boldmath$\Theta$}}

\newcommand{\diag}{{\hbox{diag}}}

\newcommand{\trace}{{\hbox{tr}}}

\newcommand{\eqdef}{\stackrel{\Delta}{=}}

\newcommand{\herm}{{\sf H}}

\newcommand{\SINR}{{\sf SINR}}
\newcommand{\SNR}{{\sf SNR}}

\usepackage{stmaryrd} 

\begin{document}

\setlength{\abovedisplayskip}{1pt}
\setlength{\belowdisplayskip}{1pt}
\setlength{\abovedisplayshortskip}{1pt}
\setlength{\belowdisplayshortskip}{1pt}

\title{Uplink-Downlink Duality and Precoding Strategies with Partial CSI in Cell-Free Wireless Networks}

\author{\IEEEauthorblockN{Fabian G\"ottsch\IEEEauthorrefmark{1},
		Noboru Osawa\IEEEauthorrefmark{2}, Takeo Ohseki\IEEEauthorrefmark{2}, Kosuke Yamazaki\IEEEauthorrefmark{2}, Giuseppe Caire\IEEEauthorrefmark{1}}
	\IEEEauthorblockA{\IEEEauthorrefmark{1}Technical University of Berlin, Germany\\
		\IEEEauthorrefmark{2}KDDI Research Inc., Japan\\
		Emails: \{fabian.goettsch, caire\}@tu-berlin.de, \{nb-oosawa, ohseki, ko-yamazaki\}@kddi-research.jp}}

\maketitle


\begin{abstract}
We consider a {\em scalable} user-centric  wireless network with dynamic cluster formation as 
defined by Bj\"ornsson and Sanguinetti. 
After having shown the importance of dominant channel subspace information for uplink (UL) pilot decontamination and having examined different UL combining schemes in our previous work, here we investigate precoding strategies for the downlink (DL). 
Distributed scalable DL precoding and power allocation methods are evaluated for different antenna distributions, user densities and 
UL pilot dimensions. We compare distributed power allocation methods to a scheme based on a particular form of UL-DL duality which is computable 
by a central processor based on the available partial channel state information. The new duality method achieves almost symmetric 
``optimistic ergodic rates'' for UL and DL while saving considerable computational complexity since the UL combining vectors are reused as DL precoders.   
\end{abstract}

\vspace{-.4cm}
\section{Introduction} 
\vspace{-.1cm}
{\em Multiuser MIMO} is a key transformative technology at the center of the last decade of theoretical research and 
practical system design, since the first information theoretical breakthrough of
Caire and Shamai \cite{caire2003achievable}, to the provisions in recent wireless standards \cite{3gpp38211}. 
A successful related concept is Marzetta's {\em massive MIMO} \cite{marzetta2010noncooperative}. 
This is based on the observation that, thanks to channel reciprocity and TDD operations,  
an arbitrarily large number $M$ of base station (BS) antennas can be trained by a finite number $K$ of user equipments (UE) 
using a finite-dimensional uplink (UL) pilot field $\tau_p \geq K$. 
Since the channel coherence block of $T$ channel uses\footnote{We define a channel coherence block as a ``tile'' of $T$ symbols in the time-frequency domain over which the fading channel can be considered constant. For the sake of conceptual simplicity, this can be identified as resource block (RB) 
of the underlying PHY protocol.}  is limited by the channel time and frequency selectivity, in a large cellular network the number of  simultaneoutly active users $K$ is generally much larger than $\tau_p$. This implies that mutually non-orthogonal pilot sequences are used in different cells 
yielding  {\em pilot contamination}, which creates a coherently combined term in the
inter-cell interference that does not vanish as $M \rightarrow \infty$ \cite{marzetta2010noncooperative}.  
More recently, a flurry of works advocating the joint processing of spatially distributed remote radio heads (RRHs) has appeared. 
This idea can be traced back to the work of Wyner \cite{wyner1994shannon}, and has been ``re-marketed''  several times under different names with slight nuances, 
such as {\em coordinate multipoint} (CoMP),  {\em cloud radio access network} (CRAN), or {\em cell-free massive MIMO}. 
An excellent recent review of this vast literature is given in \cite{9336188}. While these approaches were initially marketed as 
``eliminating inter-cell interference'',  this is unfortunately not quite true since the pilot contamination problem due to the limited UL pilot dimension persists even with full joint processing of all the RRHs. In addition, other challenges emerge. 
First, deploying a number of RRHs much larger than the number of UEs is often infeasible, especially for outdoor systems.  
Second, the joint processing of all RRHs across the network  yields a non-scalable architecture. 

In this work, we consider the realistic case of a network covering an area $A$, where the number of RRHs $L$ is less than the number of users $K$ simultaneously 
active on any given resource block (RB). However, we allow each RRH to have an $M$-antennas array, such that the total number of antenna elements $ML$ is generally larger than $K$.  We employ a scalable {\em user-centric} architecture based on dynamic cluster formation, as proposed in several recent papers
(see \cite{9336188}), such that every UE is served by a finite-size cluster of RRHs even  if the network is arbitrarily large. 
We extend our previous work \cite{goettsch2021impact} by investigating DL precoding and power allocation schemes. 
UL-DL SINR duality in cell-free wireless networks was studied in \cite{9064545} and shown to hold in terms of the Signal to Interference plus Noise Ratio (SINR) 
expression resulting from the so-called {\em use-and-then-forget (UatF) bound}, where all necessary terms to compute the ``true'' UatF bound are assumed 
to be somehow known to a central processor (CP) devoted to DL power allocation. 
Unfortunately, the UatF bound yields often overly pessimistic results in terms of achievable rate especially in the context of cell-free networks where the channel hardening phenomenon is not as pronounced as in standard cellular massive MIMO networks. Therefore, in this paper we consider the so-called ``optimistic ergodic rate'', i.e., the achievable rate when the UE receiver knows the useful signal term and and the interference plus noise power. 

Our main contribution is a new UL-DL duality in terms of SINR expressions that are actually computable by the CP using the knowledge of the large
scale fading coefficients (LSFC) and the {\em partial} CSI available at the RRHs through the cluster formation and UL pilot allocation. 
Although these SINR expressions, referred to as ``nominal SINR'' in this work, do not correspond to any achievable rate, we show  via extensive simulation that 
the UL and DL optimistic ergodic rates when the DL power is allocated through this form of duality are virtually identical. 
This ``nominal duality''  has the advantage that no DL precoding vectors need to be computed, as they are defined to be equal to the UL combing vectors. However, 
it incurs the disadvantage of centralized power allocation computation, requiring $O(K^2)$ signal coefficients to be collected at a processing unit to compute 
the duality DL powers. We compare the DL rates achieved with the proposed nominal UL-DL duality to equal power allocation (EPA), using the proposed combining methods in \cite{goettsch2021impact}. Further, we compare with two state-of-the-art distributed schemes,
 local partial zero-forcing from \cite{9069486} and local zero-forcing adapted to a system with user-centric clusters. 


\vspace{-.01cm}
\section{System model}
We refer the reader to our previous work \cite{goettsch2021impact} for a detailed system model description, and provide a summary in the following. We consider a cell-free wireless network in TDD operation mode with $L$ RRHs, each equipped with $M$ antennas, 
and $K$ single-antenna UEs. Both RRHs and UEs are distributed on a squared region on the 2-dimensional plane. 
As a result of the cluster formation scheme in \cite{goettsch2021impact} with SNR threshold $\eta$, each UE $k$ is connected to a cluster $\Cc_k \subseteq [L]=\{1,2,\dots,L\}$ of RRHs 
and each RRH $\ell$ has a set of associated UEs $\Uc_\ell \subseteq [K]$. The UE-RRH association is described by 
a bipartite graph $\Gc$ with two classes of nodes (UEs and RRHs) such that the neighborhood of UE-node $k$ is $\Cc_k$ 
and the neighborhood of RRH-node $\ell$ is $\Uc_\ell$. The set of edges of $\Gc$ is denoted by $\Ec$, i.e., $\Gc = \Gc([L], [K], \Ec)$. 
We assume OFDM modulation and that the channel in the time-frequency domain follows the
standard block-fading model \cite{marzetta2010noncooperative,9336188,9064545}. The channel vectors from UEs to RRHs are random but constant over coherence blocks of 
$T$ signal dimensions in the time-frequency domain.
The described methods are formulated for one RB, so the RB index is omitted for simplicity. 

We let $\HH \in \CC^{LM \times K}$ denote the channel matrix between all the $K$ UE antennas and all the $LM$ 
RRH antennas on a given RB, formed by $M \times 1$ blocks $\hv_{\ell,k}$ in correspondence of the $M$ antennas of RRH $\ell$ 
and UE $k$. 
We define the {\em ideal partial CSI} regime
where each RRH $\ell$ has perfect knowledge of the channel vectors $\hv_{\ell,k}$ for $k \in \Uc_\ell$. 
In this regime, the part of the channel matrix $\HH$ known  at the decentralized processing unit serving cluster $\Cc_k$ 
is denoted by $\HH(\Cc_k)$. This matrix has the same dimensions of $\HH$,  such that the $(\ell, j)$ block 
of dimension $M \times 1$  of $\HH(\Cc_k)$ is equal to $\hv_{\ell,j}$ for all  $(\ell, j) \in \Ec$, where $\ell \in \Cc_k$, and to $\zerov$ otherwise.  
Let $\Fm$ denote the $M \times M$ unitary DFT matrix with $(m,n)$-elements
$\Fm_{m,n} = \frac{e^{-j\frac{2\pi}{M} mn}}{\sqrt{M}}$ for  $m, n  = 0,1,\ldots, M-1$, and consider the angular support set $\Sc_{\ell,k} \subseteq \{0,\ldots, M-1\}$ 
obtained according to the single ring local scattering model (see \cite{adhikary2013joint}). Then, the channel between RRH $\ell$ and UE $k$ is
\begin{equation} 
	\hv_{\ell,k} = \sqrt{\frac{\beta_{\ell,k} M}{|\Sc_{\ell,k}|}}  \Fm_{\ell,k} \nuv_{\ell, k}, \label{channel_model}
	\vspace{-.14cm}
\end{equation}
where, using a Matlab-like notation, $\Fm_{\ell,k} \eqdef \Fm(: , \Sc_{\ell,k})$ denotes the tall unitary matrix obtained by selecting the columns 
of $\Fm$ corresponding to the index set $\Sc_{\ell,k}$, $\beta_{\ell,k}$ is a LSFC including
pathloss, blocking effects, and shadowing, and  $\nuv_{\ell,k}$ is an $|\Sc_{\ell,k}| \times 1$ i.i.d. Gaussian vector with components 
$\sim \Cc\Nc(0,1)$. 

\subsection{Uplink data transmission} 
The UEs transmit with the same power $P^{\rm ue}$,  and we define the system parameter $\SNR \eqdef P^{\rm ue}/N_0$, 
where $N_0$ denotes the noise power spectral density.
The received $LM \times 1$ symbol vector at the $LM$ RRHs' antennas for a single channel use of the UL is given by
\begin{equation} 
	\yy^{\rm ul} = \sqrt{\SNR} \; \HH \sss^{\rm ul}   + \zz^{\rm ul}, \label{ULchannel}
\end{equation}
where $\sss^{\rm ul} \in \CC^{K \times 1}$ is the vector
of information symbols transmitted by the UEs (zero-mean unit variance and mutually independent random variables) and 
$\zz^{\rm ul}$ is an i.i.d. noise vector with components $\sim \Cc\Nc(0,1)$.  
The goal of cluster $\Cc_k$ is to produce an effective channel observation for symbol $s^{\rm ul}_k$ 
(the $k$-th component of the vector $\sss^{\rm ul}$ from the collectively received signal at the RRHs $\ell \in \Cc_k$).  
We  define the receiver {\em unit norm} vector $\vvv_k \in \CC^{LM \times 1}$ formed by $M \times 1$ blocks
$\vv_{\ell,k} : \ell = 1, \ldots, L$, such that $\vv_{\ell,k} = \zerov$ (the identically zero vector) if $(\ell,k) \notin \Ec$. 
This reflects the fact that only the RRHs in $\Cc_k$ are involved in producing a received observation for the detection of user $k$. 
The non-zero blocks $\vv_{\ell,k} :  \ell \in \Cc_k$ contain the receiver combining vectors.
The corresponding scalar combined observation for symbol $s^{\rm ul}_k$ is given by 
\begin{eqnarray}
	r^{\rm ul}_k  & = & \vvv_k^\herm \yy^{\rm ul}. 
\end{eqnarray}
%
%
%
For simplicity, we assume that the channel decoder has perfect knowledge of the exact UL SINR value 
\begin{eqnarray} 
	\SINR^{\rm ul}_k 
& = & \frac{  |\vvv_k^\herm \hh_k|^2 }{ \SNR^{-1}  + \sum_{j \neq k} |\vvv_k^\herm \hh_j |^2 },  \label{UL-SINR-unitnorm}
\end{eqnarray}
where $\hh_k$ denotes the $k$-th column of $\HH$. The corresponding UL {\em optimistic ergodic} achievable rate is given by 
\begin{equation}
R_k^{\rm ul} = \EE [ \log ( 1 + \SINR^{\rm ul}_k ) ], \label{ergodic-rate}
\end{equation}
where the expectation is with respect to the small scale fading, while conditioning on the placement of UEs and RRH, and on the cluster formation. 

\subsection{Downlink data transmission} 
\vspace{-.001cm}
The signal corresponding to one channel use of the DL at the receiver of UE $k$ is given by 
\begin{equation} 
	y_k^{\rm dl} = \hh_k^\herm \xx  + z_k^{\rm dl},  \label{DLchannel}
\end{equation}
where the transmitted vector $\xx \in \CC^{LM \times 1}$ is formed by all the signal samples sent collectively from the RRHs. Without loss of generality
we can incorporate a common factor ${\SNR}^{-1/2}$ in the LSFCs, which is equivalent to rescaling the noise at the UEs receivers such that  
$z_k^{\rm dl} \sim \Cc\Nc(0, \SNR^{-1})$, while keeping $\beta_{\ell,k}$ for all $(\ell,k)$ identical to the UL case.
Let $\sss^{\rm dl} \in \CC^{K \times 1}$ denote the vector of information bearing symbols for the $K$ users, assumed to 
be zero mean, independent, with variance $q_k \geq 0$. 
Under a general linear precoding scheme, we have
\begin{equation} 
	\xx = \UU \sss^{\rm dl}, \label{linear-precoding}
\end{equation}
where $\UU \in \CC^{LM \times K}$ is the overall precoding matrix, formed by $M \times 1$ blocks $\uv_{\ell,k}$ such that
$\uv_{\ell,k} = \zerov$ if $(\ell, k) \notin \Ec$.  The non-zero blocks $\uv_{\ell,k} :  (\ell,k) \in \Ec$ contain the precoding vectors. Using (\ref{linear-precoding}) in (\ref{DLchannel}), we have
\begin{eqnarray}
	y^{\rm dl}_k & = & \hh_k^\herm \uu_k s^{\rm dl}_k   + \sum_{j \neq k} \hh_k^\herm \uu_j s^{\rm dl}_j  + z^{\rm dl}_k, 
\end{eqnarray}
where $\uu_k$ is the $k$-th column of $\UU$.  The resulting DL (optimistic) SINR, is given by 
\begin{eqnarray}
	\SINR^{\rm dl}_k & = & \frac{|\hh_k^\herm \uu_k|^2 q_k}{\SNR^{-1} + \sum_{j\neq k}   |\hh_k^\herm \uu_j|^2 q_j }, \label{DL-SINR} 
\end{eqnarray}
where $q_k$ is the DL transmit (Tx) power for UE $k$ and, like in the UL, the  DL achievable rate is 
$
	R_k^{\rm dl} = \EE [ \log ( 1 + \SINR^{\rm dl}_k ) ]. \label{ergodic-rate-dl}
$
Assuming that the columns of the precoder $\UU$ have unit norm, we have that the total DL Tx power collectively transmitted by the RRHs is given by 
\begin{align} 
	P^{\rm dl}_{\rm tot} &= \trace \left ( \EE [ \xx \xx^\herm ] \right ) = \trace \left ( \UU \diag(q_k : k \in [K]) \UU^\herm \right ) \\
	&= \trace \left ( \UU^\herm \UU \diag(q_k : k \in [K]) \right ) = \sum_{k=1}^K q_k.  
\end{align}
We assume the UL and DL total transmit power to be balanced, which after scaling the LSFCs with the factor $\SNR$, imposes the condition $\sum_{k=1}^K q_k = K$. 

\vspace{-.1cm}
\section{UL schemes} \label{sec:ul_perfect_csi}
\vspace{-.1cm}
In the UL, we employ a RRH cluster wise global zero-forcing (GZF) combining scheme, as well as local linear MMSE (LMMSE) with global combining. Both methods are explained in details in \cite{goettsch2021impact} and are described only briefly in the following.
\subsection{Global Zero-Forcing (GZF)}
\vspace{-.13cm}
For a given UE $k$ with cluster $\Cc_k$,  we define the set $\Uc(\Cc_k) \eqdef \bigcup_{\ell \in \Cc_k} \Uc_\ell$ of UEs served by at least one RRH in $\Cc_k$. 
Let $\hh_k(\Cc_k)$ denote the $k$-th column of $\HH(\Cc_k)$ and let $\HH_k(\Cc_k)$ denote the residual matrix after deleting the $k$-th column. 
The GZF receiver vector is obtained as follows.  Let $\overline{\hh}_k(\Cc_k) \in \CC^{|\Cc_k|M \times 1}$ and $\overline{\HH}_k(\Cc_k) \in  \CC^{|\Cc_k|M \times (|\Uc(\Cc_k)|-1)}$ the vector and matrix
obtained from $\hh_k(\Cc_k)$ and $\HH_k(\Cc_k)$, respectively, after removing all the $M$-blocks of rows corresponding to 
	RRHs $\ell \notin \Cc_k$ and all the (all-zero) columns corresponding to UEs $k' \notin \Uc(\Cc_k)$. Consider the 
	singular value decomposition 
$
		\overline{\HH}_k(\Cc_k) = \overline{\AA}_k \overline{\SSS}_k \overline{\BB}_k^\herm, 
$
	where the columns of the tall unitary matrix $\overline{\AA}_k$ form an orthonormal basis for the column subspace of $\overline{\HH}_k(\Cc_k)$, such that 
	the orthogonal projector onto the orthogonal complement of the interference subspace is given by 
$\overline{\PP}_k = \Id - \overline{\AA}_k \overline{\AA}_k^\herm$, 
and define the unit-norm vector 
	\begin{equation} 
	\overline{\vvv}_k = \overline{\PP}_k \overline{\hh}_k(\Cc_k) / \| \overline{\PP}_k \overline{\hh}_k(\Cc_k) \|. \label{gzf_combining}
	\end{equation} 
	Hence, the GZF receiver vector $\vvv_k$ is given by expanding $\overline{\vvv}_k$ by reintroducing the missing blocks of 
	all-zero $M \times 1$ vectors $\zerov$ in correspondence of the RRHs $\ell \notin \Cc_k$. 
	If $M > \tau_p$, noticing that  $|\Uc_\ell| \leq \tau_p$ (due to the cluster formation rule), we have that
	$|\Uc(\Cc_k)| \leq \tau_p |\Cc_k| < M  |\Cc_k|$.
	Therefore, $\AA_k$ defined before is effectively tall unitary and the global ZF always exists with probability 1 for random/Gaussian 
	user channel vectors. 
	
However, due to the antenna correlation introduced by the single ring local scattering model \eqref{channel_model}, it may happen that for small $M$
some UEs  $k' \in \Uc_\ell, k' \neq k$ have channels $\hh_{k'}(\Cc_k)$ co-linear with $\hh_k(\Cc_k)$ with probability 1 (with respect to the small-scale fading realizations).   
In this case, the GFZ would yield  $\vvv_k^\herm \hh_k(\Cc_k) = 0$. In order to avoid this ``zero-forcing outage'', 
we employ a simple scheme: If the cluster $\Cc_{k}$ detects a UE $k' \in \Uc(\Cc_{k})$ with a co-linear channel, it computes the GZF combining vector excluding $\hh_{k'}(\Cc_k)$, i.e., it removes the row corresponding to UE $k'$ from the matrix $\overline{\HH}_k(\Cc_k)$. 

\subsection{Local LMMSE with global combining}
\vspace{-.08cm}
In this case, each RRH $\ell$ makes use of locally computed receiving vectors $\vv_{\ell,k}$ for 
its users $k \in \Uc_\ell$. 
Let $\yv_\ell^{\rm ul}$ denote the $M \times 1$ block of $\yy^{\rm ul}$ corresponding to RRH $\ell$. 
%
	We use an LMMSE principle for the local combining vectors and distinguish between the known part of the interference, 
	i.e., the term  $\sum_{j \in \Uc_\ell : j \neq k}  \hv_{\ell,j} s_j^{\rm ul}$, and the unknown part of the interference, 
	i.e., the term  $\sum_{j \notin \Uc_\ell} \hv_{\ell,j} s_j^{\rm ul}$ in $\yv_\ell^{\rm ul}$. 
	The receiver treats the unknown part of the interference plus noise as a white vector with known variance $\sigma_\ell^2$ per component (again, see \cite{goettsch2021impact} for details). 
	Under this assumption, we have that the LMMSE receiving vector is given by 
	\begin{equation} 
		\vv_{\ell,k} = \left ( \sigma_\ell^2 \Id + \SNR \sum_{j \in \Uc_\ell} \hv_{\ell,j} \hv_{\ell,j}^\herm \right )^{-1} \hv_{\ell,k}. \label{eq:lmmse}
	\end{equation}
	The overall receiver vector is obtained by forming the vector $\overline{\vvv}_k$ by stacking the $|\Cc_k|$ blocks  of dimensions $M \times 1$ given by  $w_{\ell,k} \vv_{\ell,k}$ on top of each other and normalizing such that $\overline{\vvv}_k$ has unit norm. The combining coefficients $w_{\ell, k}$ are chosen to maximize the nominal SINR based on partial CSI at RRH $\ell$ after combining. Their computation is given in \cite{goettsch2021impact}. 
	After expanding $\overline{\vvv}_k$ to $\vvv_k$ of dimension $LM \times 1$ by inserting the all-zero blocks corresponding to the RRHs $\ell \notin \Cc_k$, the resulting  SINR is again given by (\ref{UL-SINR-unitnorm}). 
	

\section{DL precoding and power allocation schemes } \label{sec:dl_schemes}

We shall consider expressions of {\em nominal}, i.e., assumed, SINR, where the channels $\hv_{\ell,k}$ with $(\ell,k) \in \Ec$ are known, while for $(\ell,k) \notin \Ec$, only the LSFCs of their channels are known, and the spatial distribution is assumed to be isotropic. In \cite{9064545} in contrast, if RRH $\ell$ serves UE $j$ but not UE $k$, $\EE[ |  \hh^\herm_k \vvv_j |^2 ]$ is still assumed to be estimated accurately at RRH $\ell$, where $\hv_{\ell,k}^\herm \vv_{\ell,j} $ is a non-zero component of $\hh^\herm_k \vvv_j$. 

\subsection{UL-DL SINR duality for partially known channels}
Consider the UL SINR given in (\ref{UL-SINR-unitnorm}). By construction we have that $\vv_{\ell,k} = \zerov$ for all $\ell \notin \Cc_k$. Therefore, the term at the numerator
$ \theta_{k,k} = \big | \vvv_k^\herm \hh_k \big |^2 = \big| \sum_{\ell\in \Cc_k} \vv_{\ell,k}^\herm \hv_{\ell,k} \big|^2 $
contains only known channels and it can thus be used in the nominal SINR expression. 
The terms at the denominator take on the form
\begin{eqnarray}
	\theta_{j,k} & = & \left | \vvv_k^\herm \hh_j \right |^2 = \bigg | \sum_{\ell \in \Cc_k} \vv_{\ell,k}^\herm \hv_{\ell,j} \bigg |^2 \\
	& = & \bigg| \sum_{\ell \in \Cc_k \cap \Cc_j} \vv_{\ell,k}^\herm \hv_{\ell,j}  + \sum_{\ell \in \Cc_k \setminus \Cc_j} \vv_{\ell,k}^\herm \hv_{\ell,j} \bigg|^2 \label{known-unknown}
\end{eqnarray}
where for $\ell \in \Cc_k \cap \Cc_j$ the channel $\hv_{\ell,j}$ is known, while for $\ell \in \Cc_k \setminus \Cc_j$ the channel $\hv_{\ell,j}$ is not known. 
Taking the conditional expectation of the term in (\ref{known-unknown}) given all the known CSI $\Ec_k$ at $\Cc_k$, and noticing that the channels for different
$(\ell,j)$ pairs are independent and have mean zero, we find
\begin{eqnarray} 
	\Resize{1\linewidth}{\EE[ \theta_{j,k} | \Ec_k] = \bigg | \sum\limits_{\ell \in \Cc_k \cap \Cc_j} \vv_{\ell,k}^\herm \hv_{\ell,j} \bigg |^2 + 	\sum\limits_{\ell \in \Cc_k\setminus \Cc_j} \frac{\beta_{\ell,j} M}{|\Sc_{\ell,j}|} \vv_{\ell,k}^\herm \Fm_{\ell,j} \Fm_{\ell,j}^\herm \vv_{\ell,k}.  \nonumber}
\end{eqnarray}
Finally, under the isotropic assumption, we replace the actual covariance matrix of the unknown channels with
a scaled identity matrix with the same trace, i.e., we have
\begin{equation} 
	\EE[ \theta_{j,k} | \Ec_k] \approx \bigg | \sum_{\ell \in \Cc_k \cap \Cc_j} \vv_{\ell,k}^\herm \hv_{\ell,j} \bigg |^2 + 
	\sum_{\ell \in \Cc_k\setminus \Cc_j} \beta_{\ell,j} \| \vv_{\ell,k} \|^2.  \nonumber
\end{equation}
Using the fact that $\vvv_k$ is a unit-norm vector and assuming that the  $M \times 1$ blocks $\vv_{\ell,k}$ have the same norm, we can further approximate
$\| \vv_{\ell,k}\|^2 \approx \frac{1}{|\Cc_k|}$. Therefore, the resulting nominal UL SINR is given by 
\begin{eqnarray} 
	& &\SINR_k^{\rm ul-nom} \nonumber \\
	& &=  \frac{\left | \vvv_k^\herm \hh_k \right |^2}{\SNR^{-1} + \sum\limits_{j \neq k}  \Big( 
		\Big | \sum\limits_{\ell \in \Cc_k \cap \Cc_j} \vv_{\ell,k}^\herm \hv_{\ell,j} \Big |^2 + \frac{1}{|\Cc_k|} \sum\limits_{\ell \in \Cc_k\setminus \Cc_j} \beta_{\ell,j}  \Big)} \nonumber \\
	& &=  \frac{\theta_{k,k}}{\SNR^{-1} + \sum_{j\neq k} \theta_{j,k} },
\end{eqnarray}
where the coefficient $\theta_{k,k}$ is defined as before, and we redefine the coefficients $\theta_{j,k}$ for $j \neq k$ as
\begin{equation} 
	\theta_{j,k} =  
	\bigg | \sum_{\ell \in \Cc_k \cap \Cc_j} \vv_{\ell,k}^\herm \hv_{\ell,j} \bigg |^2 + \frac{1}{|\Cc_k|} \sum_{\ell \in \Cc_k\setminus \Cc_j} \beta_{\ell,j}.  \label{new-theta-jk}
\end{equation}
Notice also that the term at the denominator of the nominal UL SINR given by $\frac{1}{|\Cc_k|}  \sum_{j \neq k} \sum_{\ell \in \Cc_k\setminus \Cc_j} \beta_{\ell,j} $
is the contribution of the interference power per RRH caused by other UEs $j \neq k$ to the RRHs in $\Cc_k$, but not also in $\Cc_j$.
Next, we consider the DL SINR given in (\ref{DL-SINR}) under the assumption $\uu_k = \vvv_k$ for all $k \in [K]$. 
Again, the numerator takes on the form $\theta_{k,k} q_k$ where $\theta_{k,k}$ is the same as before and contains all known channels. 
Focusing on the terms at the denominator and taking into account that the vectors $\uu_j$ are non-zero only for the $M\times 1$ blocks corresponding to RRHs $\ell \in \Cc_j$, we have
\begin{eqnarray}
	\theta_{k,j} & = & \left | \hh_k^\herm \uu_j \right |^2 = \bigg | \sum_{\ell \in \Cc_j} \hv_{\ell,k}^\herm \uv_{\ell,j} \bigg |^2 \\
	& = & \bigg | \sum_{\ell \in \Cc_j \cap \Cc_k} \hv_{\ell,k}^\herm  \uv_{\ell,j}  + \sum_{\ell \in \Cc_j \setminus \Cc_k} \hv_{\ell,k}^\herm \uv_{\ell,j} \bigg |^2, \label{known-unknown-DL}
\end{eqnarray}
where, as before,  for $\ell \in \Cc_j \cap \Cc_k$ the channel $\hv_{\ell,k}$ is known, while for $\ell \in \Cc_j \setminus \Cc_k$ the channel $\hv_{\ell,k}$ 
is not known.  Taking the conditional expectation of the term in (\ref{known-unknown-DL}) given all the known channel state information, 
we find
\begin{equation} 
	\Resize{1\linewidth}{\EE[ \theta_{k,j} | \Ec_k] = \bigg | \sum\limits_{\ell \in \Cc_j \cap \Cc_k} \hv_{\ell,k}^\herm \uv_{\ell,j} \bigg |^2 + 
	\sum\limits_{\ell \in \Cc_j \setminus \Cc_k} \frac{\beta_{\ell,k} M}{|\Sc_{\ell,k}|} \uv_{\ell,j}^\herm \Fm_{\ell,k} \Fm_{\ell,k}^\herm \uv_{\ell,j}. \nonumber } 
\end{equation}
Using again the isotropic assumption, we replace the actual covariance matrix of the unknown channels with
a scaled identity matrix with the same trace, i.e., we have
\begingroup
\allowdisplaybreaks
\begin{eqnarray} 
	\EE[ \theta_{k,j} | \Ec_k] & \approx & 
	\bigg | \sum_{\ell \in \Cc_j \cap \Cc_k} \hv_{\ell,k}^\herm \uv_{\ell,j}  \bigg |^2 + 
	\sum_{\ell \in \Cc_j\setminus \Cc_k} \beta_{\ell,k} \| \uv_{\ell,j} \|^2 \nonumber \\
	& \approx & 
	\bigg | \sum_{\ell \in \Cc_j \cap \Cc_k} \hv_{\ell,k}^\herm \uv_{\ell,j}  \bigg |^2 + 
	\frac{1}{|\Cc_j|} \sum_{\ell \in \Cc_j\setminus \Cc_k} \beta_{\ell,k}. \nonumber
\end{eqnarray}
\endgroup
Therefore, the resulting nominal DL SINR is given by 
\begin{eqnarray} 
	& & \SINR_k^{\rm dl-nom} \nonumber \\
	& & = \Resize{.89\linewidth}
	{\frac{\left | \hh_k^\herm \uu_k  \right |^2 q_k}{\SNR^{-1} + \sum\limits_{j\neq k} \left ( 
			\Big | \sum\limits_{\ell \in \Cc_j \cap \Cc_k} \hv_{\ell,k}^\herm \uv_{\ell,j} \Big |^2 + \frac{1}{|\Cc_j|} \sum\limits_{\ell \in \Cc_j\setminus \Cc_k} \beta_{\ell,k} \right ) q_j } } \nonumber \\
	& & = \frac{\theta_{k,k} q_k}{\SNR^{-1} + \sum_{j\neq k} \theta_{k,j} q_j}, \label{DL-nom-SINR} 
\end{eqnarray}
where the coefficient $\theta_{k,k}$ is defined as before, and we redefine the coefficients $\theta_{k,j}$ for $j \neq k$ as
\begin{equation} 
	\theta_{k,j} = 
	\bigg | \sum_{\ell \in \Cc_j \cap \Cc_k} \hv_{\ell,k}^\herm \uv_{\ell,j} \bigg |^2 + \frac{1}{|\Cc_j|} \sum_{\ell \in \Cc_j\setminus \Cc_k} \beta_{\ell,k}  .
	\label{new-theta-kj}
\end{equation}
Given the symmetry of the coefficients, an UL-DL duality exists for 
the nominal SINRs. This can be used to calculate the DL Tx power allocation $\{q_k : k \in [K]\}$ that achieves DL nominal SINRs equal to the UL nominal SINRs with uniform UL Tx power per UE. 

In particular, we choose the target DL SINRs $\{\gamma_k\} \eqdef \{\SINR_k^{\rm ul-nom}\}$ for all $k$. The system of (non-linear) equations in the power allocation vector $\qv = \{q_k\}$ given by 
\[ \SINR^{\rm dl-nom}_k = \gamma_k, \;\;\; \forall \; k = 1\ldots, K \]
can be rewritten in the more convenient linear form
\begin{equation} 
	\left ( \Id  - \diag(\muv) \Thetam \right ) \qv  = \frac{1}{\SNR} \muv,  
\end{equation} 
by defining the vector $\muv$ with elements 
$
	\mu_k = \frac{\gamma_k}{(1 + \gamma_k) \theta_{k,k}} \label{mumu}
$
and the matrix $\Thetam$ with $(k,j)$ elements $\theta_{k,j}$. It can be shown that the target SINRs $\{\gamma_k\}$ are achievable 
if and only if the above system has a non-negative solution, i.e., if
\begin{equation} 
	\qv^\star  = \frac{1}{\SNR} \left ( \Id  - \diag(\muv) \Thetam \right )^{-1}  \muv \label{eq_q_duality}
\end{equation} 
is a non-negative vector and satisfies
$
\sum_{k=1}^K q^\star_k = K.
$

\subsection{Centralized ZF precoding}
We use the global ZF combining vectors computed in (\ref{gzf_combining}) as precoding vectors, i.e., $\uu_k = \vvv_k$ for all $k$, and consider two power allocation schemes. The first option is power allocation based on UL-DL duality, where the DL Tx power allocation vector $\qv$ is computed as in (\ref{eq_q_duality}). 
The second option is EPA, such that $q_k = 1$ for all UEs. 

\subsection{Local LMMSE and centralized power allocation}
The unit norm combining vector $\vvv_k$ constructed with the locally computed LMMSE combining vectors from (\ref{eq:lmmse}) is used as precoding vector $\uu_k$. The same power allocation methods are considered as for centralized ZF precoding. The UL-DL duality power allocation is computed with (\ref{eq_q_duality}) by taking the nominal UL SINRs with local LMMSE combining as target DL SINRs.

\subsection{Comparison with the state of the art: Local partial ZF}
The local partial ZF (LPZF) is adapted from \cite{9069486} to our system with user-centric clusters. Each RRH computes the precoding vectors and power allocation locally for its associated UEs. Let us consider the channel matrix between RRH $\ell$ and the associated UEs in $\Uc_\ell$, given by
\begin{equation}
	\Hm_\ell = [\hv_{\ell,k_1} \ \hv_{\ell,k_2} \dots \hv_{\ell,k_{|\Uc_\ell|}} ] \in \CC^{M \times | \Uc_\ell | },
\end{equation}
where $k_1, \dots, k_{|\Uc_\ell|}$ are the UEs in the set $\Uc_\ell$. 
In case $M \geq \tau_p \geq |\Uc_\ell|$ and $\Hm_\ell$ is a full rank matrix, local ZF (LZF) is carried out by computing the pseudoinverse
\begin{eqnarray}
	\Hm_\ell^{+} = \Hm_\ell \left ( \Hm_\ell^{\herm} \Hm_\ell \right )^{-1} \label{local_zf} 
\end{eqnarray} 
of $\Hm_\ell$. Then, the LPZF precoding vector $\uv_{\ell, k}$ is the normalized column of $\Hm_\ell^{+}$ corresponding to user $k \in \Uc_\ell$.

In the other case that $M \leq |\Uc_\ell|$, or that $\Hm_\ell$ is rank-deficient ($\tau_p \geq |\Uc_\ell|$ due to clustering), the RRH chooses from $\Uc_\ell$ the UEs $\Uc_\ell^{\rm ZF}$ with the largest channel gains (at most $M$) whose channels are linearly independent, and thus form a full-rank matrix $\Hm_\ell^{\rm ZF}$. The precoding vectors of UEs in $\Uc_\ell^{\rm ZF}$ are computed by ZF as in (\ref{local_zf}) and normalized to unit norm. For the remaining UEs $k \in \Uc_\ell^{\rm MRT}$, normalized MRT is employed, i.e., 
\begin{equation}
	\uv_{\ell,k} = \frac{\hv_{\ell,k}}{\| \hv_{\ell,k} \|}, \ \forall k \in \Uc_\ell^{\rm MRT},
\end{equation}
where $\Uc_\ell^{\rm ZF} \bigcap \Uc_\ell^{\rm MRT} = \emptyset$ and $\Uc_\ell^{\rm ZF} \bigcup \Uc_\ell^{\rm MRT} = \Uc_\ell$.

For both cases, the RRHs compute the Tx power for each UE locally. Two simple schemes are considered: EPA with \vspace{-.05cm}
\begin{eqnarray}
	q_{\ell, k} = \frac{P^{\rm RRH}}{| \Uc_\ell |}, \ \forall k \in \Uc_\ell , \label{eq:epa}
\end{eqnarray}
and proportional power allocation (PPA) with regard to the LSFCs such that \vspace{-.05cm}
\begin{eqnarray}
	q_{\ell, k} = P^{\rm RRH} \frac{\beta_{\ell,k}}{ \sum_{j \in \Uc_\ell} \beta_{\ell, j} }, \ \forall k \in \Uc_\ell , \label{eq:ppa}
\end{eqnarray}
where $q_{\ell, k}$ and $P^{\rm RRH}$ denote the transmit power allocated at RRH $\ell$ to UE $k$ and the DL power budget at each RRH, respectively. For $k \notin \Uc_\ell$, we have $q_{\ell, k} = 0$. The SINR with distributed DL power allocation then becomes
\begin{eqnarray}
	\SINR^{\rm dl-dist}_k = \frac{ \sum_{\ell \in \Cc_k} |\hv_{\ell,k}^\herm \uv_{\ell,k}|^2 q_{\ell,k}}{\SNR^{-1} + \sum_{j\neq k}  \sum_{\ell \in \Cc_j} |\hv_{\ell,k}^\herm \uv_{\ell,j}|^2 q_{\ell,j} } \label{DL-dist} 
\end{eqnarray}
and $	R_k^{\rm dl-dist} = \EE [ \log ( 1 + \SINR^{\rm dl-dist}_k ) ]$.
For a fair comparison with the centralized power allocation schemes, we require $P^{\rm RRH} = K/L$, such that the sum DL Tx power is the same for all schemes.

\vspace{-.12cm}
\section{UL channel estimation}
In practice, ideal partial CSI is not available and the channels $\{\hv_{\ell,k} : (\ell,k) \in \Ec\}$ must be estimated from UL pilots. Thanks to channel reciprocity in the TDD mode, the estimates can be used for both UL combining and DL precoding.
We assume that $\tau_p$ signal dimensions per RB are dedicated to UL pilots (see \cite{3gpp38211}), and define a codebook of $\tau_p$ orthogonal pilot 
sequences. 
The pilot field received at RRH $\ell$ is given by the $M \times \tau_p$ matrix of received symbols
$\Ym_\ell^{\rm pilot} = \sum_{i=1}^K \hv_{\ell,i} \phiv_{t_i}^\herm + \Zm_\ell^{\rm pilot}$, 
where $\phiv_{t_i}$ denotes the pilot vector of dimension $\tau_p$ used by UE $i$ in the current RB, with  
total energy $\| \phiv_{t_i} \|^2 = \tau_p \SNR$.  
Assuming that the subspace information $\Fm_{\ell,k}$ of all $k \in \Uc_\ell$ is known, 
for each UE $k \in \Uc_\ell$, RRH $\ell$ despreads $\Ym_\ell^{\rm pilot}$ with $\phiv_{t_k}$, and projects the despreaded signal onto the subspace spanned by the columns of $\Fm_{\ell,k}$, yielding the channel estimate
\begin{align} 
	\widehat{\hv}^{\rm sp}_{\ell,k} & =  \frac{1}{\tau_p \SNR} \Fm_{\ell,k}\Fm_{\ell,k}^\herm \Ym^{\rm pilot}_\ell \phiv_{t_k}  \\
	& =  {\hv_{\ell,k}  + \sum\limits_{i\neq k : t_i = t_k } 
		\Fm_{\ell,k}\Fm_{\ell,k}^\herm \hv_{\ell,i}  + \Fm_{\ell,k}\Fm_{\ell,k}^\herm \widetilde{\zv}_{t_k,\ell}} ,  \label{chest}
\end{align} 
where $\widetilde{\zv}_{t_k,\ell}$ is i.i.d. with components $\Cc\Nc(0, \frac{1}{\tau_p\SNR})$. The second term in (\ref{chest}) contains the remaining pilot contamination from UEs $i \neq k$ using the same pilot $t_k$ (see \cite{goettsch2021impact} for details).


Note that the previously described UL and DL schemes, and the computation for UL-DL duality with ideal partial CSI can be carried out with channel estimates by replacing 
the ideal partial CSI $\{ \hv_{\ell,k} : (\ell,k) \in \Ec\}$ with the estimates $\{\widehat{\hv}^{\rm sp}_{\ell,k} : (\ell,k) \in \Ec\}$. 
\vspace{-.02cm}
\section{Simulations and concluding remarks} 
\vspace{-.02cm}
\begin{figure}[t]
	\centerline{\includegraphics[width=.5\linewidth]{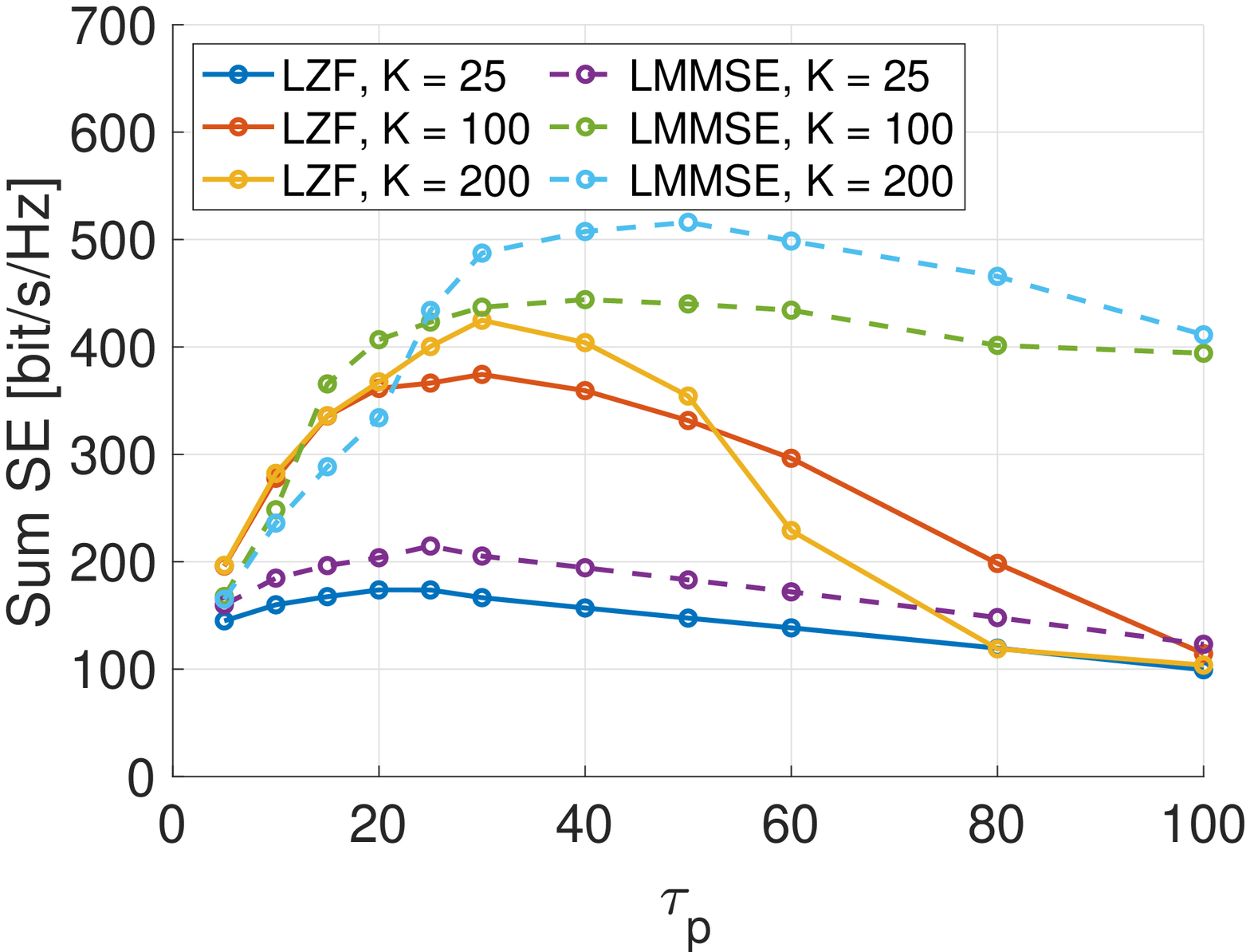} \hspace{.0cm} \includegraphics[width=.5\linewidth]{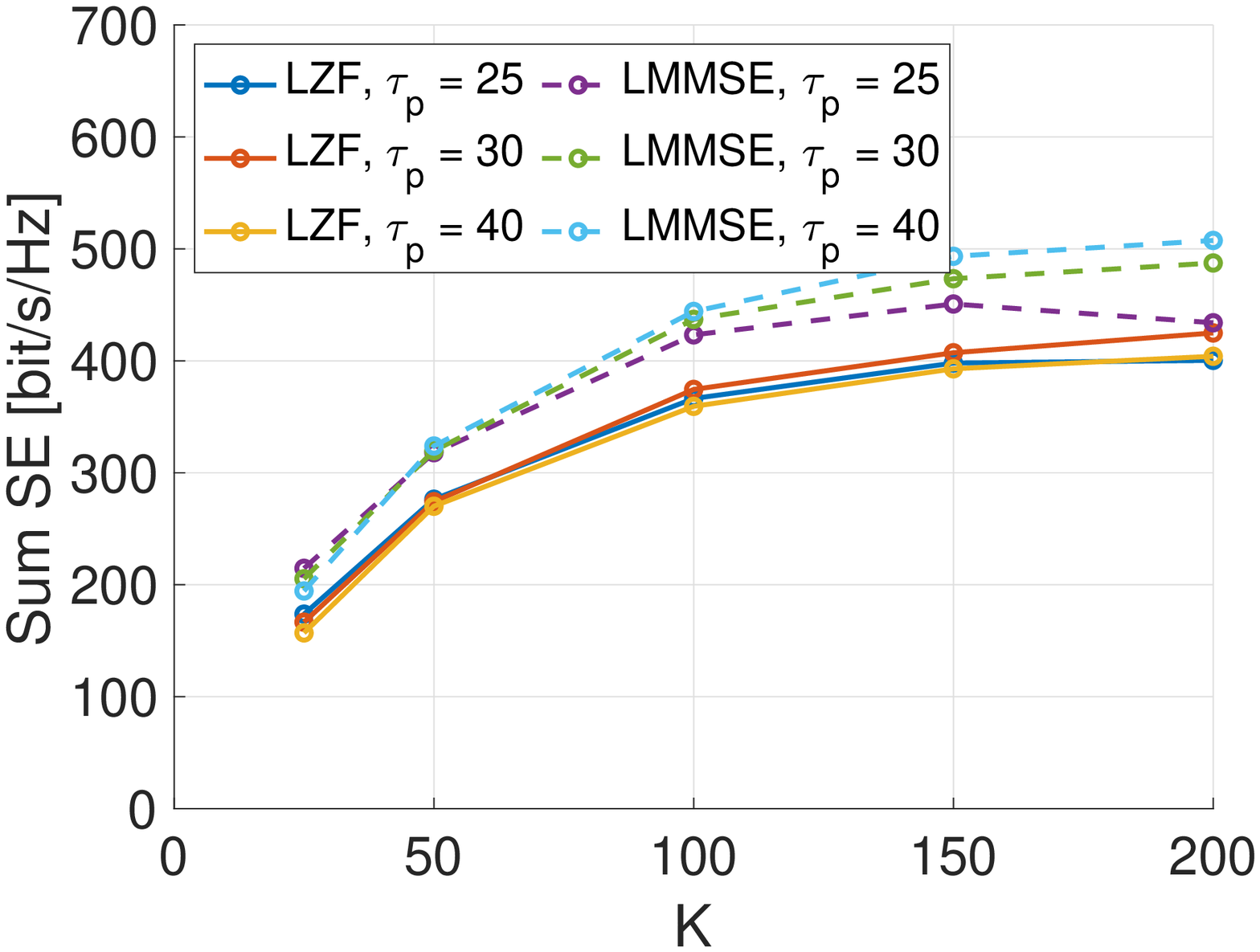}}
	\vspace{-.3cm}
	\caption{Sum DL spectral efficiency for different $K$ and $\tau_p$, where $L=10$ and $M=64$. LMMSE with power allocation from duality, LZF with PPA.}
	\vspace{-.2cm}
	\label{fig:compare_K_taup_L10}
\end{figure}
\begin{figure}[t]
	\centerline{\includegraphics[width=.5\linewidth]{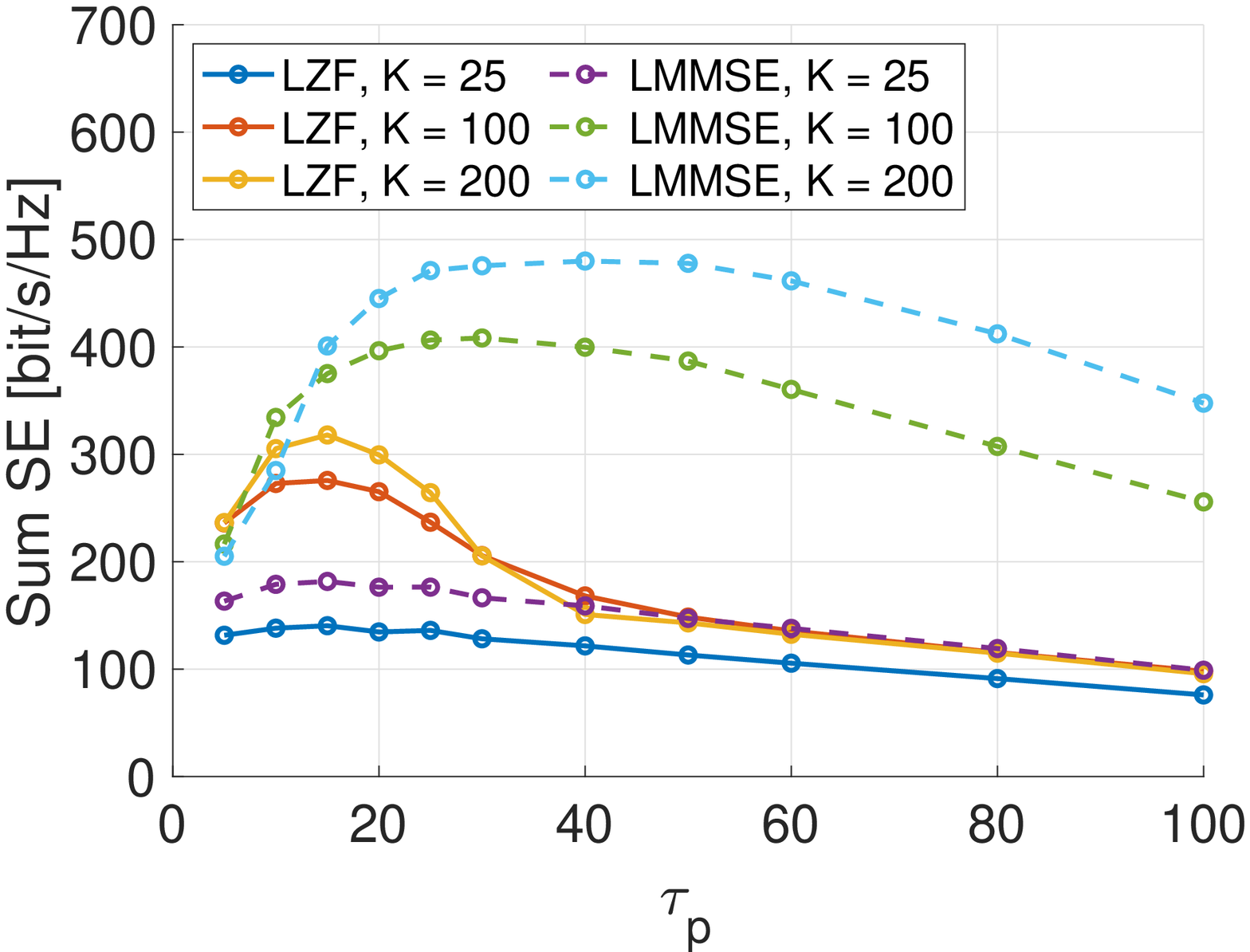} \hspace{.0cm} \includegraphics[width=.5\linewidth]{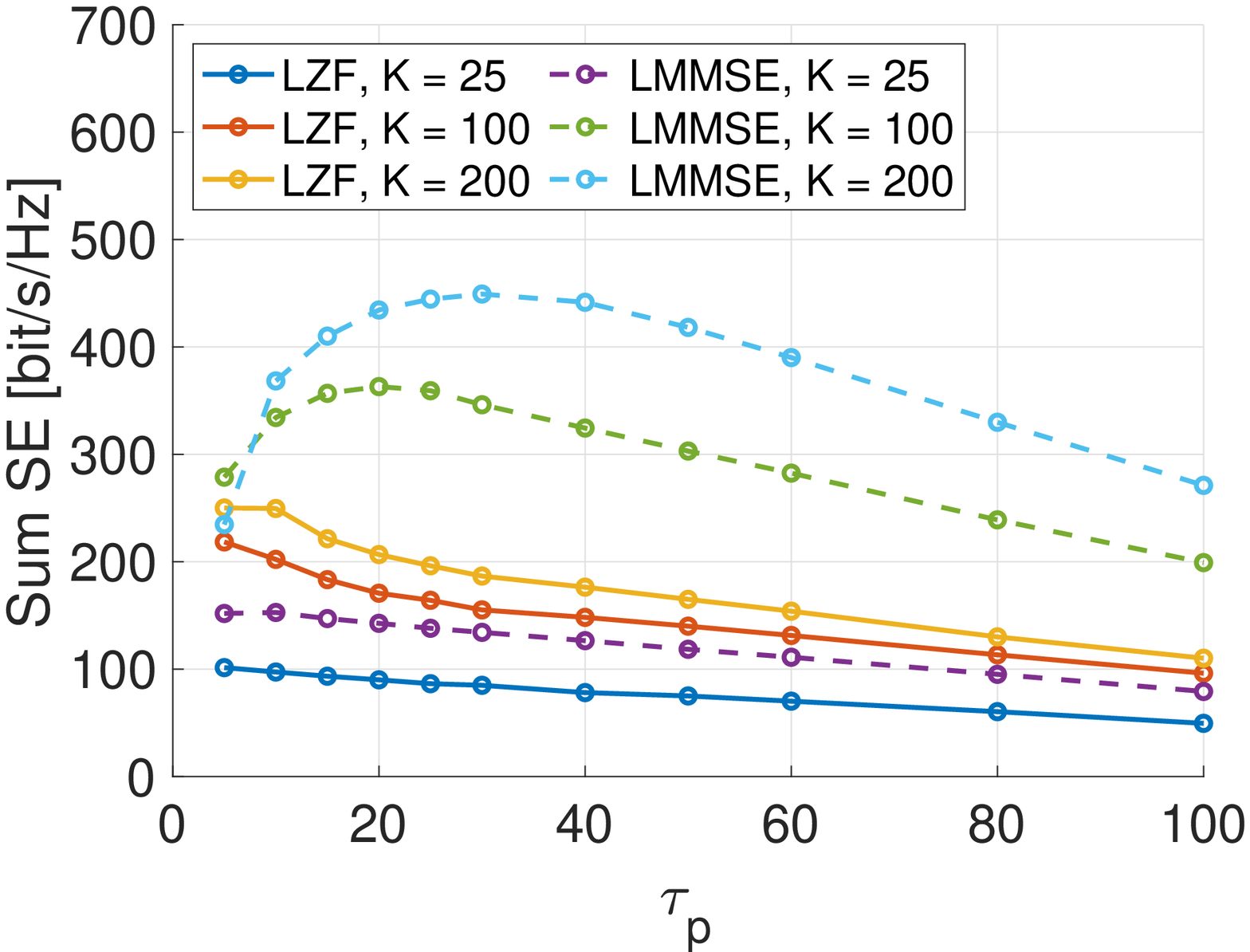}}
	\vspace{-.3cm}
	\caption{Sum DL spectral efficiency for different $K$ and $\tau_p$, where $L=20$, $M=32$ (left) and $L=40$, $M=16$ (right). LMMSE with power allocation from duality, LZF with PPA.}
	\vspace{-.59cm}
	\label{fig:compare_K_taup_L20}
\end{figure}


In our simulations, we consider a square coverage area of $A = 225 \times 225$ square meters with a torus topology to avoid boundary effects. 
The LSFCs are given  according to the 3GPP urban microcell pathloss model from \cite{3gpp38901}. 
With $N_0=-96 \text{dBm}$, the UL power $P^{\rm ue}$ is chosen such that $\bar{\beta} M \SNR = 1$ (i.e., 0 dB), when the expected pathloss $\bar{\beta}$ with respect to LOS and NLOS is calculated for distance $3 d_L$, where $d_L = 2 \sqrt{\frac{A}{\pi L}}$ is the diameter of a disk of area equal to $A/L$.
We consider RBs of dimension $T = 200$ symbols. 
The UL (same for DL) spectral efficiency (SE) for UE $k$ is given by  
\begin{equation}
	{\rm SE}^{\rm ul}_k =  (1 - \tau_p/T) R_k^{\rm ul}.
\end{equation}
The angular support $\Sc_{\ell,k}$ contains the DFT quantized angles (multiples of $2\pi/M$) falling inside an interval of length $\Delta$ placed symmetrically around the direction joining UE $k$ and RRH $\ell$. We use $\Delta = \pi/8$ and the maximum cluster size $Q=10$ (RRHs serving one UE) in the simulations. The SNR threshold $\eta=1$ makes sure that an RRH-UE association can only be established, when $\beta_{\ell,k} \geq \frac{\eta}{M \SNR } $.
For each set of parameters we generated 50 independent layouts (random uniform placement of RRHs and UEs), 
and for each  layout we computed the expectation in (\ref{ergodic-rate}) by Monte Carlo averaging with respect to the channel vectors. 
In all figures, the results with subspace projection channel estimates are shown. 
We compare the DL schemes GZF and LMMSE to LPZF \cite{9069486} (see Section \ref{sec:dl_schemes}) and local ZF (LZF) as benchmarks. The LZF precoding vectors at RRH $\ell$ are computed with (\ref{local_zf}) after selecting at most $M$ UEs out of the set $\Uc_\ell$ with the largest LSFCs and linearly independent channels. The other $| \Uc_\ell | - M$ UEs are declared in outage by RRH $\ell$. For the active UEs, the precoding vectors are normalized and power allocation follows (\ref{eq:epa}) or (\ref{eq:ppa}). 

We start our evaluation by comparing the sum SE for different $K$ and $\tau_p$ in a system with $L=10$ and $M=64$. Fig. \ref{fig:compare_K_taup_L10} shows that in most cases LMMSE outperforms LZF, and that for more UEs in the system, we need larger $\tau_p$ to maximize the sum SE, as each RRH can serve up to $\tau_p$ UEs.
For increasing $K$ and fixed $\tau_p$, the sum SE grows until $K=150$ for all $\tau_p$ and both LZF and LMMSE. Further increasing $K$ however leads to smaller improvements, and in case of LMMSE and $\tau_p = 25$ to a drop in the sum SE, as the per-cluster information becomes too partial.
Fig. \ref{fig:compare_K_taup_L20} shows a similar behavior of the sum SE for $L=20$, $M=32$ and $L=40$, $M=16$, where the best results are achieved with smaller $\tau_p$ compared to $L=10$.
Comparing the sum SE for the different levels of distribution of the antennas in the system, i.e., the ratio $L/M$ while keeping $LM$ fixed, the largest sum SE for both LZF and LMMSE is achieved with $L=10$, $M=64$.

\begin{figure}[t]
	\centerline{\includegraphics[width=.5\linewidth]{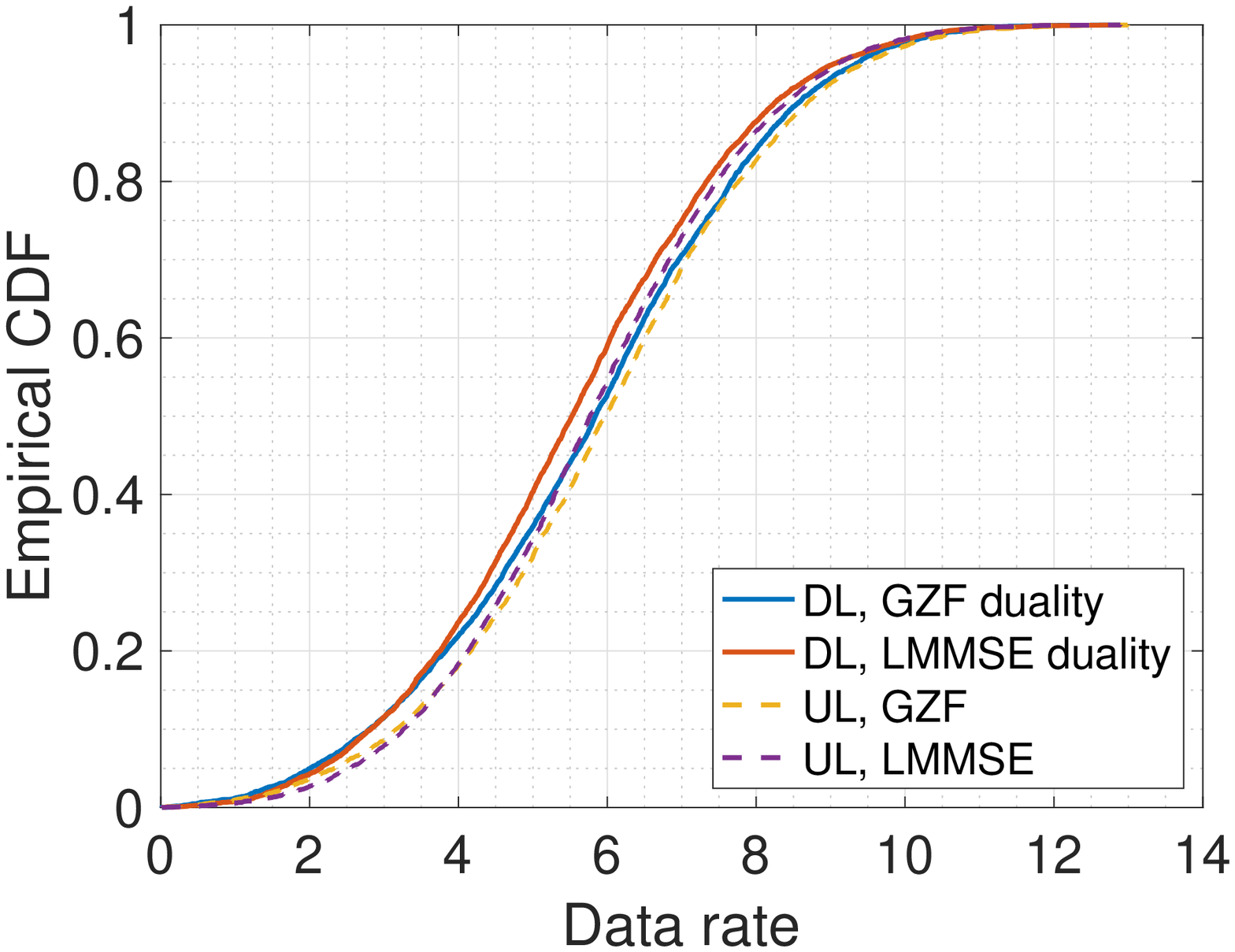} \hspace{.0cm} \includegraphics[width=.5\linewidth]{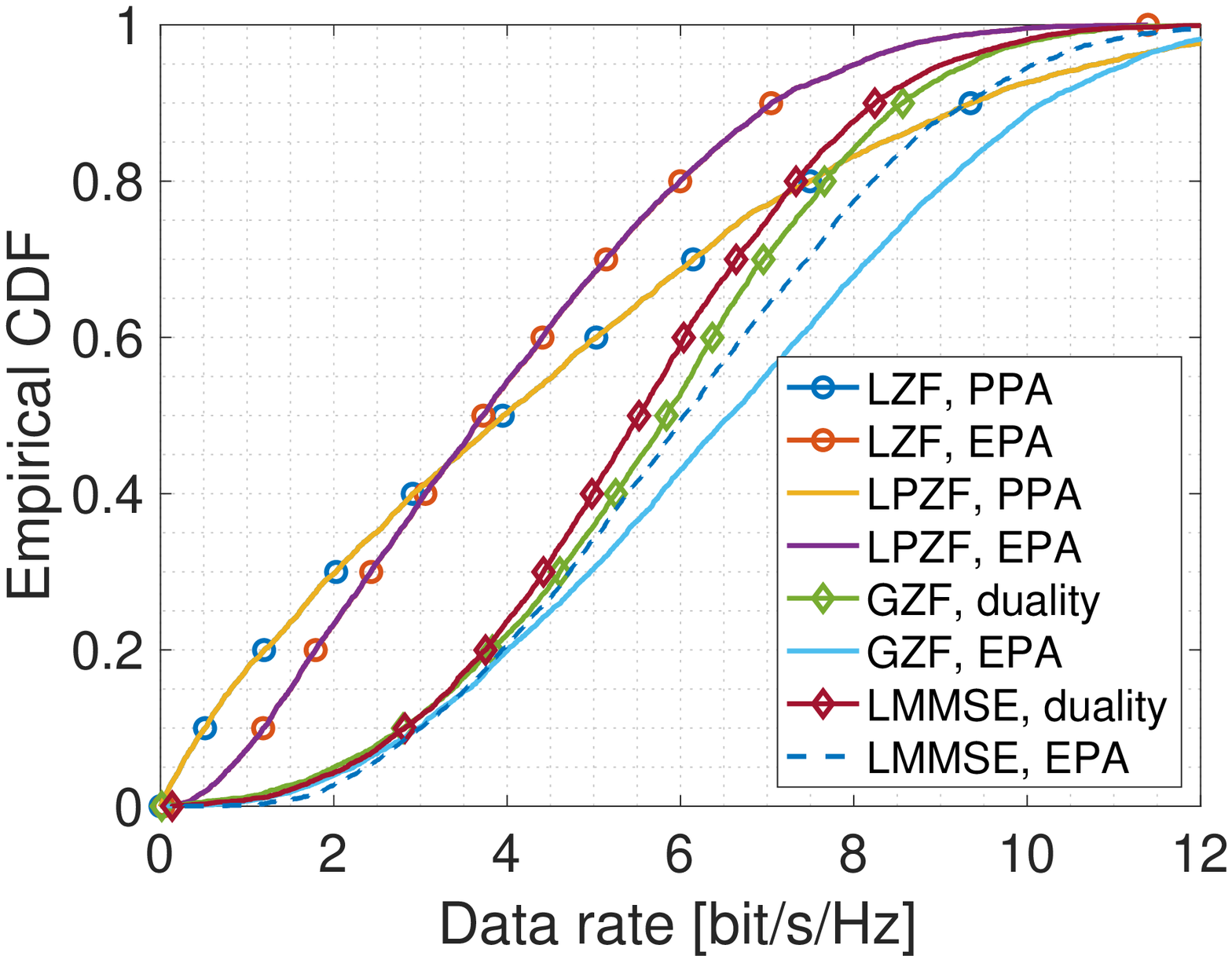}}
	\vspace{-0.3cm}\caption{Empirical cdf of the data rates per UE for UL-DL duality (left Fig.) and of the DL data rates (right Fig.), where $L=10$, $M=64$ and $\tau_p=40$.}
	\vspace{-.6cm}
	\label{fig:rates_cdf_L10}
\end{figure}

\begin{figure}[t]
	\centerline{\includegraphics[width=.5\linewidth]{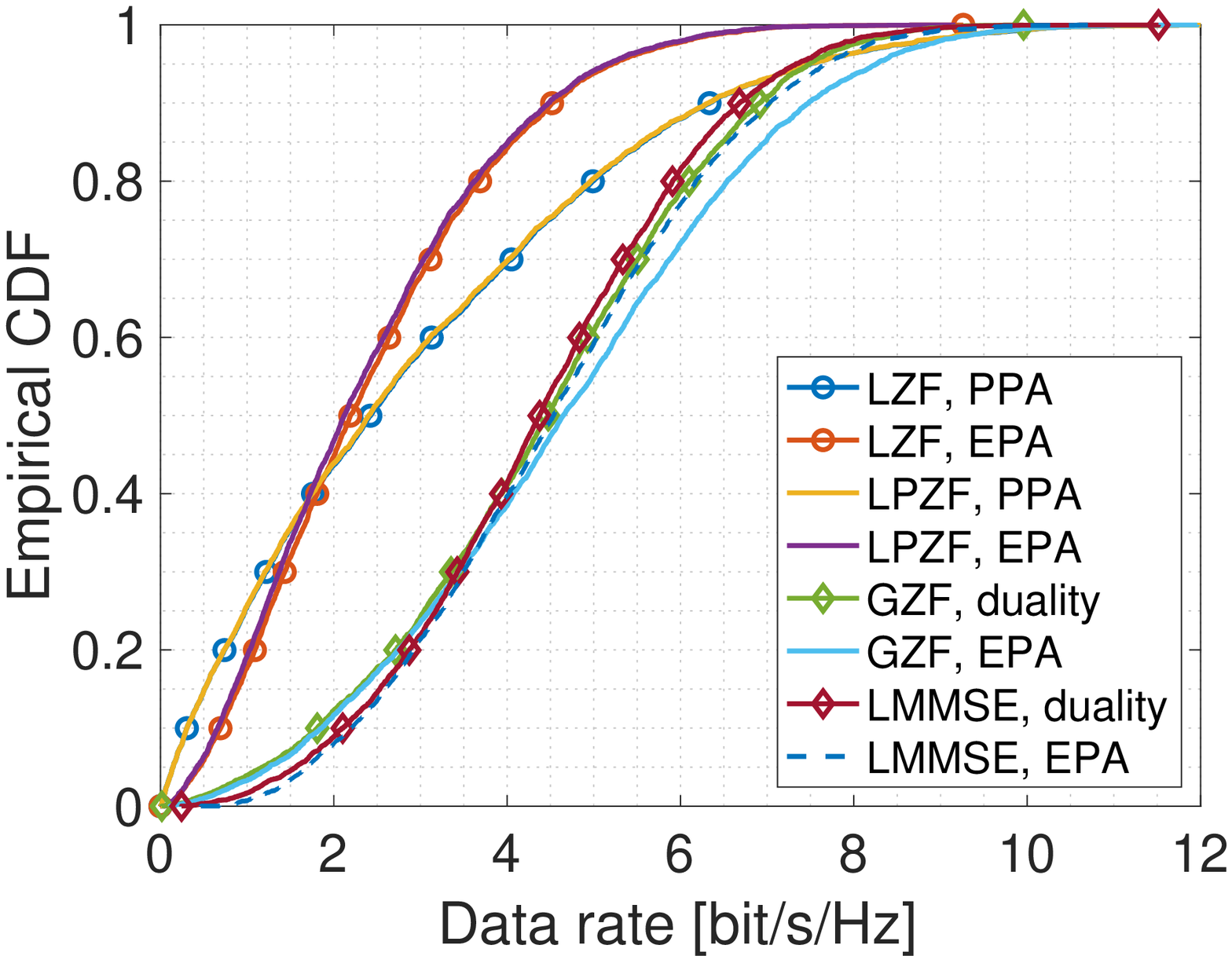} \hspace{.0cm} \includegraphics[width=.5\linewidth]{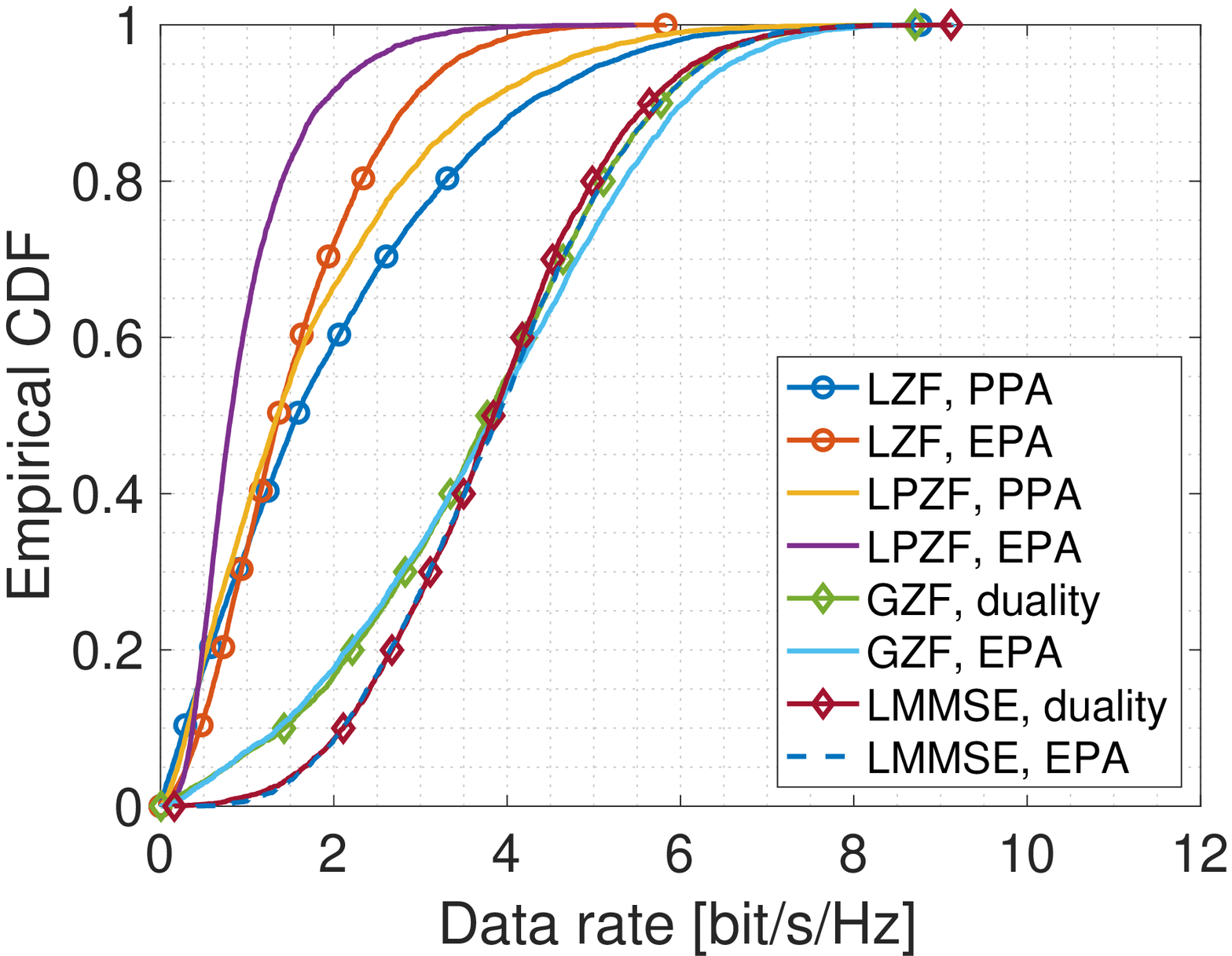}}
	\vspace{-0.3cm}\caption{Empirical cdf of the DL data rates per UE, where $L=20$, $M=32$, $\tau_p=20$ (left) and $L=40$, $M=16$, $\tau_p=15$ (right).}
	\vspace{-.5cm}
	\label{fig:rates_cdf_L20_L40}
\end{figure}

We now look at the distribution of the DL rates per UE. We set $K=100$ and $\tau_p$ for each combination of $L$ and $M$, such that both LZF and LMMSE yield a relatively large sum SE, i.e., for $L = \{10, 20, 40\}$ we have $\tau_p = \{ 40, 20, 15 \}$, respectively.
Fig. \ref{fig:rates_cdf_L10} shows that the proposed UL-DL duality method yields almost symmetric effective ergodic rates for the UL and DL. In the DL for $L=10$, the GZF and LMMSE methods outpferform the local schemes at the UEs with low rates. The UEs with the largest 30\% of rates have approximately the same performance for GZF, LMMSE, and LPZF/LZF with PPA.
Fig. \ref{fig:rates_cdf_L20_L40} shows that the gap between GZF/LMMSE and the local schemes grows for larger $L$, as there are more possible interfering signals from other RRHs and the maximum beamforming gain $M$ decreases. 

The LMMSE scheme yields a fairer distribution of the data rates with respect to GZF, while GZF  outperforms LMMSE at the UEs with high rates.
For both LMMSE and GZF, EPA leads to higher data rates as power allocation from duality and the best performance is achieved when $L=10$. Note that EPA has the additional advantage that the $K^2 + K$ parameters $\theta_{j,k}$ and $\gamma_k$ for all $j,k \in [K]$ do not need to be collected for the computation of the DL Tx powers, which makes these schemes {\em scalable}.

The local methods yield higher rates for smaller $L$ and larger $M$, like GZF and LMMSE. Both the PPA and EPA yield similar results for $L=10$, but as $L$ is increased, the gap between them grows, where PPA outperforms EPA for most UEs. EPA leads to a fairer distribution of rates in all scenarios though.
While LZF and LPZF obtain basically the same DL rates for  $M=32$ and $M=64$, LZF outperforms the LPZF scheme for $M=16$. The same performance for $M=32$ and $M=64$ is explained by the fact that it very rarely occurs that the channels of two or more UEs are co-linear for large $M$. Also $M>\tau_p$, so ZF is possible. In that case, all UEs can be served with ZF precoding by the LPZF method and it becomes equal to LZF. 
In case of $M=16$ though, the probability that a RRH serves two or more UEs in the same angular region is larger than for $M=32$ and $M=64$. One of these UEs, say $k^{\rm ZF} \in\Uc_\ell^{\rm ZF}$, with co-linear channels  is served with ZF, the other $k' \in \Uc_\ell^{\rm MRT}$ with MRT in the LPZF scheme. While the signal for $k^{\rm ZF}$ will not cause any interference to the UEs in $\Uc_\ell$, the signal for the UEs $k' \in \Uc_\ell^{\rm MRT}$ will. Employing LZF is the better choice, as no additional interference is caused by serving UEs with MRT.
\vspace{-.41cm}
\section{Conclusions}
For the simulated system, we can conclude that a more concentrated antenna distribution such as $L=10$ and $M=64$ leads to a larger sum SE for both per-cluster and local DL precoding schemes. UL-DL duality yields almost symmetric data rates, and in the DL we can achieve the largest data rates with the cluster-wise schemes using EPA, which has the additional advantage that less cooperation among RRHs is required.

\bibliography{IEEEabrv,wcnc-paper}

\end{document}